\documentclass[prl,amsmath,amssymb,amsfonts,twocolumn,nofootinbib]{revtex4-1}
\usepackage{bm,graphicx,mathrsfs}

\usepackage{graphicx}
\usepackage{epsfig}
\usepackage{amsmath,bbm}
\usepackage{amsfonts,amssymb}
\usepackage{times}
\usepackage{color}

\newcommand{\1}{\mathbbm{1}}
\newcommand{\id}{{\rm id}}
\newcommand{\cL}{{\cal L}}

\newcommand{\cO}{\mathcal{O}}

\newcommand{\cP}{{\cal P}}

\newcommand{\cC}{{\cal C}}

\newcommand{\tr}{{\rm tr}}

\newcommand{\be}{\begin{equation*}}
\newcommand{\ee}{\end{equation*}}
\newcommand{\bea}{\begin{eqnarray*}}
\newcommand{\eea}{\end{eqnarray*}}

\def\>{\rangle}
\def\<{\langle}

\newcommand{\ket}[1]{|#1\rangle}
\newcommand{\bra}[1]{\langle#1|}

\definecolor{james}{rgb}{1,.6,0}

\newcommand{\qed}{}

\def\Proof{\noindent{\it Proof.}}
\def\proof{\Proof}
\def\qed{\leavevmode\unskip\penalty9999 \hbox{}\nobreak\hfill
     \quad\hbox{\leavevmode  \hbox to.77778em{%
               \hfil\vrule   \vbox to.675em%
               {\hrule width.6em\vfil\hrule}\vrule\hfil}}
     \par\vskip3pt}
     
\begin{document}

\newtheorem{theorem}{Theorem}
\newtheorem{lemma}[theorem]{Lemma}
\newtheorem{corollary}[theorem]{Corollary}
\newtheorem{proposition}[theorem]{Proposition}
\newtheorem{definition}[theorem]{Definition}
\newtheorem{example}[theorem]{Example}
\newtheorem{conjecture}[theorem]{Conjecture}

\title{Precisely timing dissipative quantum information processing}

\author{M.\ J.\ Kastoryano$^{1,2}$}
\author{M.\ M.\ Wolf$^{3}$}
\author{J.\ Eisert$^{1}$}
\affiliation{$^1$ Qmio Group, Dahlem Center for Complex Quantum Systems, Freie Universit\"at Berlin, 14195 Berlin, Germany\\
$^2$ Niels Bohr Institute, Blegdamsvej 17, DK-2100 Copenhagen \O, Denmark\\ $^3$ Department of Mathematics, Technische Universit\"at M\"unchen, 85748 Garching, Germany
}
\date{\today}

\begin{abstract} 
Dissipative engineering constitutes a framework within which quantum information processing protocols are powered by system-environment interaction rather than by unitary dynamics alone. This framework embraces noise as a resource, and consequently, offers a number of advantages compared to one based on unitary dynamics alone, e.g., that the protocols are typically independent of the initial state of the system. 
However, the time independent nature of this scheme makes it difficult to imagine precisely timed sequential operations, conditional measurements or error correction. In this work, we provide a path around these challenges, by introducing basic dissipative gadgets which allow us to precisely
initiate, trigger and time dissipative operations, while keeping the system Liouvillian time-independent. These
gadgets open up novel perspectives for thinking of timed dissipative quantum information processing.
As an example, we sketch how measurement based computation can be simulated in the dissipative setting.
\end{abstract}
\maketitle



One of the main goals in quantum information science and in quantum technologies in general 
is to understand the information processing power available within the framework of quantum mechanics. Several quantum computational models have been proposed and analyzed, and each has its strengths and weaknesses. Every such model will suffer severely from the problem of decoherence due to noise from a practically inevitable environment, so that a lot of effort has been invested in developing methods for isolating systems from their destructive environment. 
In such a picture, quantum noise and dissipation is seen as necessarily being detrimental for coherent
quantum state manipulation.
Recently, however,  a new avenue has been suggested for addressing this issue: actively exploiting the dissipation into the environment. In a number of recent studies, it has been shown, both experimentally \cite{Krauter,Barreiro} and theoretically \cite{VWC,Diehl,Cooling,Kraus,Critical,DissMem,DissDist,KRS,MPC} that several of the basic tasks in quantum information science (quantum computation \cite{VWC}, state preparation \cite{VWC,Diehl,Kraus,KRS,MPC}, distillation \cite{DissDist}, 
storage \cite{DissMem}, and
cooling \cite{Cooling}) can be performed by engineering the system environment coupling, instead of isolating a subsystem and performing coherent controlled unitary dynamics on it. 

The most counterintuitive application of such a paradigm might be 
dissipative quantum computing (DQC) \cite{VWC}; a model of quantum computation in which one assumes a system-environment interaction described by a Markovian master equation, where the computation is encoded in the Lindblad operators and the outcome of the computation is encoded in the unique stationary state of the open system. 
The main benefit with this approach (that under appropriate conditions the unique stationary state is reached rapidly starting from any initial state) appears to imply a potentially severe disadvantage: that operations cannot be performed sequentially in time. Steps of a procedure cannot be conditioned on previous steps, it is not clear how to ``stop preparing'' a state, and it is far from clear how to incorporate error correction into any such scheme.

In this work, we open up new perspectives for dissipative quantum information processing by introducing 
 and analyzing a number of dissipative ``gadgets''.
They can be combined to act as time triggers for various dissipative operations; 
i.e., they allow a certain dissipative process to start (or stop) acting after a very specific point in time:
They serve as convenient ``clocks'' in a framework with time-independent Liouvillians.
The functioning of these tools is based on a counterintuitive property of classical Markov chains called the ``cutoff phenomenon''
\cite{Diaconis}, which can be lifted up into the
quantum setting \cite{cutoffpaper}. We also show how such gadgets can be used
in order to ``translate'' any scheme involving unitary dynamics, measurements and conditional dynamics into
a time independent dissipative setting. We then outline how these time triggers can be used to perform  dissipative measurement based quantum computation \cite{MBQC}. 

Although universal quantum computation can be performed dissipatively without recourse to timing of operations, or initializations, we show here that using these simple and intuitive gadgets, one can obtain results which may be difficult to obtain otherwise. Indeed, the timer gadgets can be used to show that DQC is universal also for geometrically local interactions ($6$-local if embedded in a 3D lattice). In the original proof of DQC \cite{VWC}, the interactions were made 5-local, but in a way which cannot be embedded in a lattice geometry.  


We show that timed dissipation-driven
quantum information processing is indeed possible. Needless to say, 
there is a lot of room to reduce the significant 
overhead involved in order to make the schemes more suitable for experiments. 

The paper is set up as follows. We first introduce an \textit{initialization gadget} which allows to prepare a given state for a finite amount of time. Next, we combine the initialization gadget with a special behavior from Markov chain mixing
 to construct a rudimentary \textit{timer gadget}, which allows to trigger a dissipative operation at a specific point in time. Rigorous error estimates for these two gadgets are given in the Appendix. We go on to outline an application of these gadgets: a {\it dissipative one-way quantum computation} (D1WQC) scheme.

\begin{figure}[th]
\includegraphics[scale=0.36]{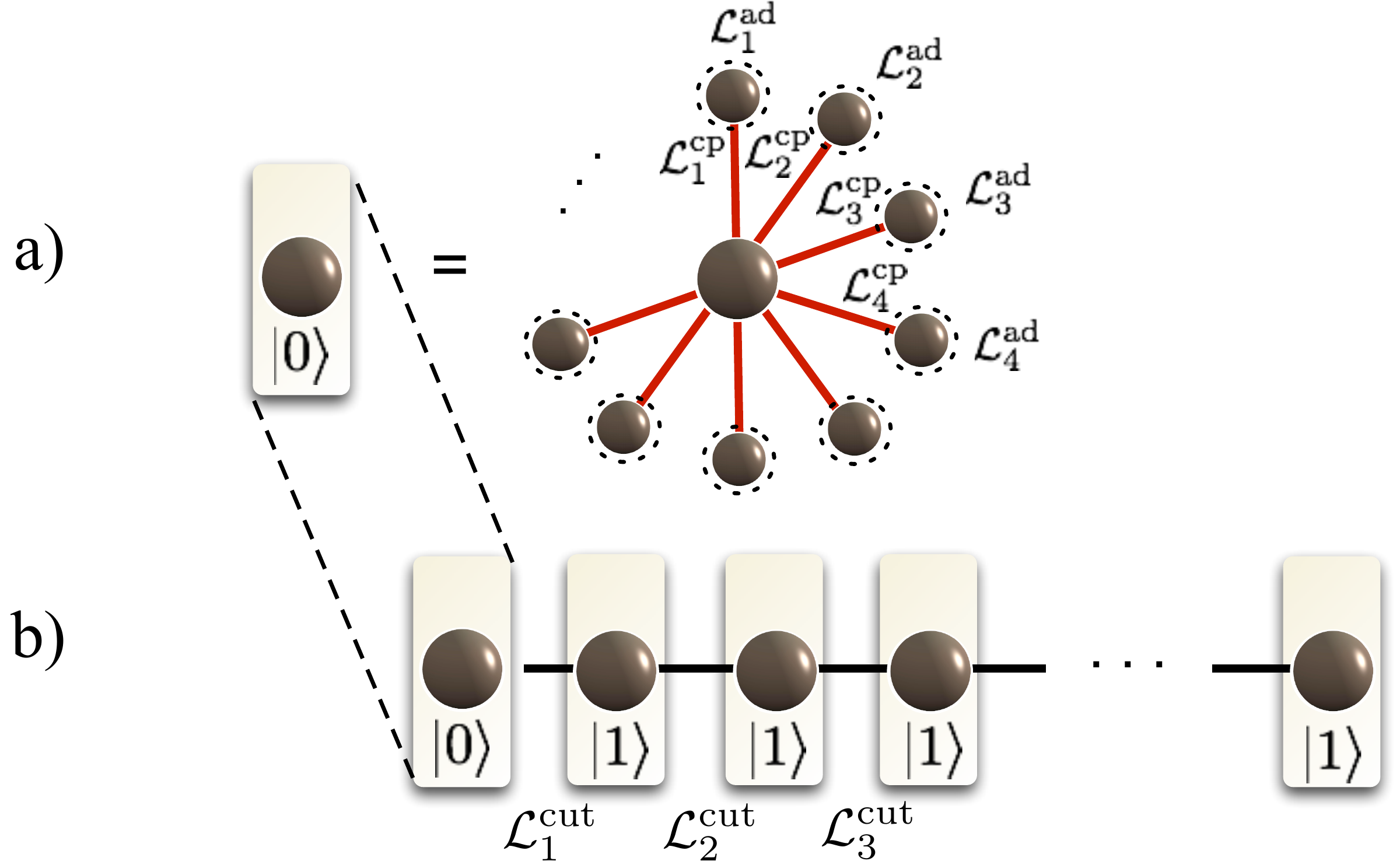}\label{Star}
\caption{a) Depiction of an initialization gadget. The central circle depicts the target qubit, the outer circles depict the auxiliary qubits. The red lines connecting the timer qubit to the auxiliary qubits illustrate the conditional state preparation local Liouvillians $\cL^{\rm cp}_j$, and the dotted circles around the auxiliary qubits illustrate unconditional amplitude damping Liouvillians $\cL^{\rm ad}_j$; i.e. $\cL^{\rm ini}=\cL^{\rm cp}+\cL^{\rm ad} $.
b) Depiction of the linear timer gadget. The circles represent the timer qubits, the lines connecting the timer qubits are the local Liouvillians $\cL^{\rm cut}_j$, $j=1,\dots, N-1$, and the boxes illustrate the initialization devices from a).}
\end{figure}

{\it Basic building blocks.}
The basic conceptual building blocks for the dissipative gadgets considered here are: an idea for
initialization, and a particular mixing behavior called the ``cutoff phenomenon''. 
Throughout, we will assume that the open system dynamics are properly modeled by a Markovian master equation
with a local Liouvillian, so that
\be	
	\frac{d}{dt}\rho = \sum_j  \biggl( L_j \rho L_j^\dagger-\frac{1}{2}\{L_j^\dagger L_j, \rho\}\biggr),
\ee
where each Lindblad operator $L_j$ has a local support which is independent of the system size. 

{\it The initialization gadget.} 
For initialization to be successful, the qubit must (i) be prepared in the desired state in a time which scales at most polynomially in the gadget size $M$ while the error decreases exponentially with $M$, and (ii) the state preparation must stop acting after a time which is specified by the desired dissipative process. We show how to initialize a single qubit in the $\ket{0}$ state, 
with a preparation time logarithmic in the gadget size. 
For this, consider a star shaped graph as in Fig.\ 1 a), and label the central qubit as $c$ and the auxiliary qubits as $j=1,\dots,M$. 
The central qubit -- the one we want to initialize -- is surrounded by $M$ auxiliary qubits. The Liouvillian of the gadget can be written $\cL^{\rm ini}=\cL^{\rm ad}+\cL^{\rm cp}$, where $\cL^{\rm ad}$ represents a collection of amplitude damping channels acting independently on the $M$ auxiliary qubits with rate $\omega$, 
\begin{equation}
	L^{\rm ad}_k=\sqrt{\omega}\ket{0_k}\bra{1_k}, 
\end{equation}
for $k=1,\dots,M$, and $\cL^{\rm cp}$ represents the preparation of the central qubit in state $\ket{0_c}\bra{0_c}$, at a rate $\Gamma$, conditioned on the auxiliary qubits being in the state $\ket{1_k}\bra{1_k}$:
\begin{equation}
	L^{\rm cp}_k=\sqrt{\Gamma}\ket{0_c}\bra{1_c} \otimes \ket{1_k}\bra{1_k}. 
\end{equation}
Intuitively, after a time $t_M=\cO(\log{M}/\omega)$, the $M$ auxiliary qubits will all have relaxed
to $\ket{0}$ to within exponentially small error, so that the state preparation essentially stops acting. We set out to show that for all but an exponentially small fraction of input states, the central qubit will be in the desired state $\ket{0_c}$ after the time $t_M$  with an error that is exponentially small in $M$.  Importantly, we
point out that the dynamics of the initializer gadget acts exclusively on the diagonal elements of the system's density matrix.  This implies that no off-diagonal element contributes to the mixing time analysis, and hence we can assume that the initialization process is essentially classical. 

\begin{theorem}[Initialization] Let $\rho$ be an arbitrary input state, and let $\cL^{\rm ini}$ be the Liouvillian for the initialization gadget. 
If there exist $\delta,c>0$ and a subset $S\subset\{1,\dots, M\}$,  with $|S|=  c M$
such that 
$\bra{1_j}\rho_j\ket{1_j}>\delta$ for all $j\in S$. 
Then for any $\varepsilon>0$, there exists a $\tau = \cO(\log{(M)})$ such that for all $t>\tau$,
\be	
	\bra{1_c}\rho_c(t)\ket{1_c}\leq M e^{-\mu M}+\varepsilon, 
\ee 
where  $\rho_j$ is the restriction of $\rho$ to subsystem $j$, and $\rho_c(t)=\tr_{\rm aux}{e^{t\cL^{\rm ini}}(\rho)}$ is the partial trace over the auxiliary qubits, and $\mu$ is some positive constant which depends on $\{c, \delta, \omega,\Gamma\}$. 
\end{theorem}

The precise form of $\mu$  is given in the Appendix along with a proof. The theorem states that if a sufficient number of auxiliary bits have a non-zero overlap with the $\ket{1}$ state, then the stationary state of the central qubit will be exponentially close to $\ket{0_c}$. It also states that the system equilibrates in a time $\cO(\log{(M)})$. It becomes clear in the proof that the assumptions on the initial state can be further weakened, without affecting the result too severely.

The construction of the initialization gadget can be extended in several ways. The first convenient generalization is to initialize a given Lindblad operator rather than a state. This can be done simply by initializing a state, say $\ket{0}_c$ and then conditionally attaching a Lindlbald operator $L$ to $\ket{0}_c$. That way, $L$ only starts acting when $\ket{0}_c$ has been properly initialized. A second generalization is to extend the geometry of the initializer to graphs other than the star. In particular, if the initializer is constructed as a tree, with the intended initialized state as the stem, then the initialization time can be engineered to be any polynomial in $N$. This is very useful when one wants to initialize a process which only mixes in a time polynomial in $N$. Furthermore, we point out that the general tree structure for the initialization gadget also allows us to embed it in a 3D lattice with nearest neighbor interactions \cite{commentQDots}.


{\it The timer gadget.} We will now show how to trigger a dissipative operation at a particular point in time, and by concatenation, how to perform sequential dissipative operations. Intuitively, one might be tempted to introduce timescales by simply imposing different decay rates $\gamma_1\ll \gamma_2\ll \dots \ll \gamma_N$ before each desired dissipative operation $\cL_j$. But it becomes evident that either the convergence time or the error of such a process will scale poorly in $N$. A way around this problem is to further exploit  the ``cutoff phenomenon'' \cite{Diaconis}.  A dissipative process is said to exhibit cutoff-type behavior  \cite{cutoffpaper}, if there exists an initial state which does not converge to stationarity for a long time, and converges exponentially fast (as a function of the number of qubits) after a specific time. 

We provide perhaps the simplest example of a timer gadget: a line of $N$ qubits such that the last one exhibits a cutoff from $\ket{1_N}$ to $\ket{0_N}$ 
in a time $\cO(N)$. Each qubit of the timer must first be initialized in the state which will lead the chain to exhibit a cutoff. As depicted in Fig.\ 1 b), the timer gadget consists of a main line of $N$ ``timer" qubits, connected by $(N-1)$ local Liouvillians $\cL^{\rm cut}$ with Lindblad operators: 
$L^{\rm cut}_j=\sqrt{\gamma}\ket{0_j}\bra{0_j}\otimes\ket{0_{j+1}}\bra{1_{j+1}}$, 
for $j=1,\dots,N-1$. The local Liouvillian $\cL^{\rm cut}_j$  acts as an amplitude damping channel with rate $\gamma$ on qubit $j+1$ if qubit $j$ is in $\ket{0_j}$ and acts trivially otherwise. 

The initialization gadget guarantees that we start in the desired state vector $\ket{\phi_0}:= \ket{0}\otimes 
\ket{1}^{\otimes (N-1)}$; i.e., the first qubit is in $\ket{0}$ and all of the other qubits are in $\ket{1}$. The qubit which will act as a trigger is the last one, so we want to estimate how long it will take before the $N$'th qubit is in $\ket{0}$, assuming proper initialization. This can be calculated explicitly. Let $\tr_{N-1}$ be the partial trace over the first $N-1$ qubits, then the following proposition proves cutoff behavior of the timer gadget in arguably the strongest possible sense.

\begin{theorem}[Timing]\label{cutoffProp}
Let $N$ be the number of timer qubits, then for $t_N(x)=(N+x\sqrt{N})/\gamma$,
\be 
	\frac{1}{2}
	\| \tr_{N-1}(e^{t_N(x)\cL^{\rm cut}}(\phi_0)) -|0_N\rangle\langle 0_N| \|_1 = 1-\Phi(x)+\theta(1/\sqrt{N}),
\ee
where $\Phi(x)=\int_{-\infty}^x e^{-t^2/2} dt/{\sqrt{2\pi}}$ is the cumulative normal distribution function. 
\end{theorem}

We note that for any real $c>0$, it immediately follows from Theorem \ref{cutoffProp} that 
\bea 	
	\lim_{N\rightarrow \infty}  \bra{0_N}\tr_{N-1}[e^{ct_N(0)\cL^{\rm cut}}(\phi_0)]\ket{0_N}&=&0, \,\, {\rm for}\, c<1,\\ 
	\lim_{N\rightarrow \infty}  \bra{0_N}\tr_{N-1}[e^{ct_N(0)\cL^{\rm cut}}(\phi_0)]\ket{0_N}&=&1,\,\, {\rm for}\, c>1,
\eea which is the definition of a cutoff given in Ref.\ \cite{cutoffpaper}. However, Theorem \ref{cutoffProp} makes a  stronger statement yet. It gives us a cutoff time ($t_N=N/\gamma$), as well as the accurate cutoff window ($\Delta t_N=\sqrt{N}$), and the deviation away from the cutoff time in the form of the cumulative distribution function $\Phi$. This last piece of information is especially useful as it allows us to estimate the inaccuracy of the timer, and analyze the error propagation for multiple uses of the gadget.
 
It follows that if the state of the lattice is properly initialized to $\ket{\phi_0}$, then an operation conditioned on the last qubit will only start taking place after a time of order $t_N=N/\gamma$, while it is exponentially suppressed before that. The trigger will operate on a time scale of order $\sqrt{N}$, and the error in the triggering will be given by the tail of a normal distribution. In practice this means that a polynomial number of dissipative operations can be triggered each in linear time, with a linear time interval between them, while accurately triggering each operation with exponential precision. Rigorous error analysis is given in the Appendix. Finally, we note that the above linear timer gadget displays three dissipative time scales, given by: $\gamma$ the rate decay of the timer Lindblad operators, $\Gamma$ the decay rate of the initialization Lindblad operators, $\omega$ the decay rate of the amplitude damping operators on the auxiliary qubits of the initialization gadget. All of these time scales are constant in the size and in the number of the gadgets, so we can choose them in such a way that they clearly separate active dissipative processes. For optimal functionality we then want $\gamma \ll\omega \ll\Gamma$, so that the timer Lindblad operators effectively do not start acting before $\phi_0$ is initialized. We also emphasize that the timer gadget analyzed above is in no way unique, and several other models could be better suited to given situations \cite{commentRingGadget}.  We now turn to applications.

{\it Dissipative one-way quantum computation:} We sketch how to perform D1WQC using the timer and initialization gadgets. We need to be able to perform the following three main steps: (i) initialize a two dimensional lattice in a cluster state \cite{MBQC}, (ii) perform conditional projective measurements \textit{in a time ordered fashion} on the cluster state which simulate a universal gate set (here CNOT and arbitrary single qubit rotation) and (iii) propagate the classical boolean phase information from one measurement to another. All three of these steps have to be performed purely dissipatively, in such a way that the stationary state is \text{essentially} unique, and that it is reached in a time which scales polynomially in the system size. It turns out that the obvious bottlenecks in the program - initialization of the cluster state and timing of operations - are solved by the timer and initialization gadgets introduced above. It can also be seen that, by construction, every Lindblad operator is geometrically local.

Indeed, as shown in Ref.\ \cite{cutoffpaper}, a graph state \cite{GS} of $N$ qubits can be prepared dissipatively in a time which scales at most as $\cO(\log{N})$; using $5$-local Liouvillians in case of a 2D-lattice. Moreover, the state can be initialized by conditionally controlling each Lindblad operator of the graph state preparation with an initialization gadget. In particular, we can use the star geometry above, as the graph state can be prepared in time $\cO(\log{N})$.

Projective measurements can also be very naturally cast into a dissipative setting.  Let $\{\ket{\xi_k}\}$ be the orthogonal basis associated with a given projective measurement on subsystem $s$, and let $r$ denote the classical registry on which the measurement is stored, then the master equation describing the measurement is
\be  
	\cL^{\rm meas}(\rho) = \sum_k \bra{\xi_k}\rho_s\ket{\xi_k}\1_s\otimes\ket{\xi_k}\bra{\xi_k}-\rho .
\ee
Note in particular, that measurement is on the same footing as logical gates in this model, and hence, all resources are treated equally. 

In D1WQC it is also necessary to perform measurements conditional on the previous measurement outcomes. This can be performed by keeping track of the pseudo-measurements on a classical information bus which is updated by conditioning. This bus only needs to keep track of two bits per measurement and can be fed into future measurements by conditioning. This can be performed locally on a three-rail bus.  Here, each step in the D1WQC is triggered by a global clock (timer gadget).
In Ref.\ \cite{MBQC} it was shown how to perform CNOT and one-qubit rotations in the 1WQC model. The CNOT gate can be performed by simultaneous measurements on nine qubits with two input and two output qubits. Arbitrary one qubit measurements can be performed on a line of five qubits, but the measurements have to be performed sequentially with feedback. We refer to Ref.\ \cite{MBQC} for a full description of these schemes. 

Rather than performing a full analysis of the dissipative versions of these schemes, which is beyond the scope of this article, we will give a detailed analysis of a subroutine, which already contains essentially all of the main ingredients of this model: qubit state transfer along a line. Consider a 1D chain of odd-$n$ spins, where we want to transfer the state of the first spin to the last spin by directed measurements. We assume that the initial state is already encoded in the cluster state at the input. In the standard setting this is done by first preparing the chain in the state $\ket{\varphi_{\rm in}}\otimes\ket{+}^{\otimes (n-1)}$, where $\ket{\pm}:=(\ket{0}\pm\ket{1})/\sqrt{2}$, 
and then applying an entangling unitary operation $S_j=\ket{0_j}\bra{0_j}+\ket{1_j}\bra{1_j}\otimes\sigma^z_{j+1}$ onto the state. The cluster state with $\ket{\varphi_{\rm in}}$ encoded in the input will be denoted $\ket{\varphi_{\rm in}}_\cC:=S\ket{\varphi_{\rm in}}$. We will not discuss how to initialize the input state dissipatively, but the generalization should be obvious.

Now, in the standard state transfer by measurement, one would measure qubits $1$ through $n-1$ in the $\sigma_x$ basis and obtain measurement outcomes $\{s_1, \dots, s_{n-1}\}$ where $s_j\in\{0,1\}$, and $s_j=0$ corresponds to the $\ket{+_j}$ projection and $s_j=1$ corresponds to the $\ket{-_j}$ projection. The final state vector after 
$n-1$ measurements is $ \ket{\phi_{\rm out}}=\ket{s_1}\otimes\dots\otimes \ket{s_{n-1}}\otimes\ket{\varphi_{\rm out}}$. 
It is not difficult to see that $\ket{\varphi_{\rm out}}=U^{\rm st}_\Sigma\ket{\varphi_{\rm in}}$, where $U^{\rm st}_\Sigma=\sigma_x^{s_1+s_3+\dots+s_{n-2}}\sigma_z^{s_2+\dots+s_{n-1}}$. The sum over measurement outcomes should be interpreted as being ${\rm mod} (2)$. 
In order to recover the original state one needs to perform the inverse unitary $(U^{\rm st}_\Sigma)^\dag$ onto $\ket{\varphi_{\rm out}}$, or equivalently, 
if one is only interested in the measurement outcomes of $\ket{\varphi_{\rm in}}$ after the state transfer, one can simply reinterpret the results based on the 
measurement outcomes $\{s_j\}$. 
In order to illustrate the analogous dissipative construction, we show here how to perform the transfer of a state along a three qubit chain. The full treatment 
of the $n$ qubit state transfer is provided in the Appendix.

Consider a chain of three qubits, labelled $\{1,2,3\}$, together with two auxiliary qubits, labelled $\{4,5\}$. As in the standard measurement based state transfer, we prepare the cluster state of three qubits with $\ket{\varphi_{\rm in}}$ encoded in the first; i.e. $\ket{\varphi_{\rm in}}_\cC$. We will then act on the five qubits of the system with two maps sequentially; i.e. act with $e^{t\cL^{A}}$ until equilibrium is reached, and then act with $e^{t\cL^B}$. Both of these semigroups can be seen to mix rapidely, as they are conditional depolarizing semigroups with few conditioners. The sequential application of the maps can be performed by attaching each to a timer qubit which prescribes a specific starting time to each map. The Liouvillians $\cL^A$ and $\cL^B$ are given by the following Lindblad operators,
$ L^{A}_{1,j}= \sqrt{\omega}\ket{+_j}\bra{+_j}\otimes\ket{0_{j+3}}\bra{1_{j+3}}$,
$L^{A}_{2,j}= \sqrt{\omega}\ket{-_{j}}\bra{-_{j}}\otimes\ket{1_{j+3}}\bra{0_{j+3}}$, where $j=\{1,2\}$. The second Liouvillian has Lindblad operators
$ L^B_1= \sqrt{\omega}\sigma^z_3\otimes\ket{0,0}\bra{0,1}$, 
$L^B_2= \sqrt{\omega}\sigma^x_3\otimes\ket{0,0}\bra{1,0}$, 
$L^B_3= \sqrt{\omega}\sigma^z_3\sigma^x_3\otimes\ket{0,0}\bra{1,1}$, where the bras and kets with zeros and ones always refer to the auxiliary qubits. 
These maps will do the following: $e^{t\cL^A}$ takes $\ket{\varphi_{\rm in}}_\cC$ to an even superposition of 
\bea &\ket{+,+}\otimes\ket{\varphi_{\rm in}}\otimes\ket{0,0},\,
&\ket{+,-}\otimes(\sigma_z\ket{\varphi_{\rm in}})\otimes\ket{0,1},\\
&\ket{-,+}\otimes(\sigma_x\ket{\varphi_{\rm in}})\otimes\ket{1,0},\,
&\ket{-,-}\otimes(\sigma_x\sigma_z\ket{\varphi_{\rm in}})\otimes\ket{1,1}.\eea
In other words, it performs a measurement on qubits one and two but keeps a record of each outcome (on the classical registry $\{4,5\}$). The second map $e^{t\cL^B}$ acts as the unitary correction $U^{\rm st}_\Sigma$ by rotating the superposition of states above onto the vector
$\ket{\phi_{\rm out}}\bra{\phi_{\rm out}}=\1_{1,2}\otimes\ket{\varphi_{\rm in}}\bra{\varphi_{\rm in}}\otimes\ket{0,0}\bra{0,0}$ , from which $\ket{\varphi_{\rm in}}$ can be read off directly. 

The CNOT and single qubit rotation are performed in much the same way with a slightly more involved classical bus for measurement and feedback (see the Appendix for the $n$-chain qubit state transfer). Likewise, concatenation of elementary gates can be handled efficiently, as the classical bus only ever has to keep two bits of information per updated qubit.

{\it Summary and outlook.}
In this work, we have introduced a basic toolbox for timing dissipative operations with a time-independent Liouvillian. We have shown that dissipative operations can be performed sequentially with only a polynomial overhead in qubits or in Lindblad operators. We also show how to initialize (i.e., switch on and off) dissipative operations, which then leads to full dissipative state initialization. We have provided the necessary error analysis to show that these gadgets can be applied sequentially while remaining efficient. 
Finally, we outlined how these gadgets can be used to perform a dissipative version of one-way computation, holding the promise of geometrically local dissipative computation. The gadgets proposed here give rise to
a versatile toolbox for constructing novel schemes of dissipative quantum information processing, opening up avenues
for dissipative engineering. 
Finally, we mention a very promising applications for our gadgets: dissipative error correction, where the interplay between active and passive operations become critical. A fully passive error correction scheme remains elusive,  but clues suggest that dissipative engineering might be the right framework (see Appendix). It is the hope that 
the new perspectives opened up here by introducing the possibility of sequential schemes including measurement and processing trigger further research in this
direction.

{\it Acknowledgements.} We would like to thank C.\ Gogolin, F.\ Pastawski, 
and M.\ Kliesch for discussions and the EU (Q-ESSENCE, QUEVADIS), the AvH,
the EURYI, the ERC (TAQ), and the BMBF (QuOReP, CQC) for support.

\begin{appendix}

\section*{Appendix}
In this section, we prove Theorem 1, 
and we analyze an instance of a physically reasonable input state for the initialization gadget. 
To start with, we introduce some notation and state some basic observations which will be useful in the following. First note that $\cL^{\rm ini}$ commutes with both the projection into computational basis and any permutation of the $M$ outer qubits. So if we want to compute the overlap $\tr[A e^{t \cL^{\rm ini}}(\rho)]$ w.r.t. any $A$ which has the same symmetries, we can replace $\rho$ by its symmetrization which has the form 
\be\sum_{m=0}^M\sum_{a=0}^1\lambda_k^a\varphi_k^a. 
\ee
Here, the $\lambda_k^a$'a are probabilities and $\varphi_k^a=\ket{a_c}\bra{a_c}\otimes P_k$, where $P_k$ is the normalized projection onto the set of all states of the outer $M$ qubits with exactly $k$ excitations, i.e., $k$ qubits in $\ket{1}$. Without loss of generality, we can always restrict ourselves to the abelian algebra generated by the $\varphi_k^a$'s. In this way the problem becomes entirely classical. We will denote the conditional expectation, i.e., the projection onto this algebra, by $\cP$.

\subsection*{Proof of Theorem 1}
\proof{
We observe first that 
\be 
\bra{1_c}\rho_c(t)\ket{1_c}=1-\bra{0_c}\rho_{c,D}(t)\ket{0_c}
\leq\frac12 \|\rho_D(t)-\varphi_0^0\|_1
\ee where $\rho_D(t)=\cP\circ e^{t\cL^{\rm ini}}(\rho)$, $\rho_{c,D}(t)$ is the corresponding reduced density matrix of the central qubit,
and the stated inequality uses monotonicity of the trace norm under the partial trace.
Now, the stationary states of $e^{t\cL^{\rm ini}}$ are of the form 
\be
	\rho_\beta\otimes \ket{0} \bra{0}^{\otimes M} , 
\ee
where $\rho_\beta=\beta\ket{0_c}\bra{0_c}+(1-\beta)\ket{1_c}\bra{1_c}$ and $\beta\in[0,1]$. 
Therefore, consider the following two projections onto stationary states,
\bea T_\varphi(\rho)&=&(1-\bra{\varphi^1_0}\rho\ket{\varphi^1_0})\ket{\varphi^0_0}\bra{\varphi^0_0}+\bra{\varphi^1_0}\rho\ket{\varphi^1_0}\ket{\varphi^1_0}\bra{\varphi^1_0},\\
T_\infty(\rho)&=&\lim_{t\rightarrow\infty}e^{t\cL^{\rm ini}}(\rho),\eea 
and observe that $T_\varphi \circ T_\infty=T_\infty$.
Then, using the triangle inequality for the trace norm, 
\bea 
	&&\|e^{t\cL^{\rm ini}}(\rho_D)-\varphi^0_0\|_1\\
	&&\leq \| (T_\varphi\circ e^{t\cL^{\rm ini}}) (\rho_D)-\varphi^0_0\|_1
	+\|(T_\varphi \circ e^{t\cL^{\rm ini}}-e^{t\cL^{\rm ini}})(\rho_D)\|_1.
\eea The second term can be bounded as 
\bea &&\|(T_\varphi \circ e^{t\cL^{\rm ini}}-e^{t\cL^{\rm ini}})(\rho_D)\|_1\\&&\leq\|(T_\varphi \circ e^{t\cL^{\rm ini}} - T_\varphi \circ T_\infty)(\rho_D)\|_1+\|(e^{t\cL^{\rm ini}}-T_\infty)(\rho_D)\|_1\\&&\leq 2\|(e^{t\cL^{\rm ini}}-T_\infty)(\rho_D)\|_1\leq 6 Me^{-t\omega}\eea with the last inequality following from Lemma \ref{Lem:added}.

The remaining term can be bounded by using Lemma \ref{initk}, and writing $\rho_D$ in terms of the basis $\varphi^{(0,1)}_k$. Given our assumption that there exists a $\delta>0$ and a subset $S\subset\{1,\dots, M\}$, $|S|=  c M$ for some $c>0$,
such that $\bra{1_j}\rho_j\ket{1_j}>\delta$ for all $j\in S$, we deduce that the probability of measuring all zeros in the input state is upper bounded by $(1-\delta)^{c M}$. Similarly, the probability of measuring $k$ ones ($k\leq c M$) and all the rest zeros is upper bounded by $(1-\delta)^{(c M-k)}$. 
Thus, writing $\rho_D=\sum_{k=0}^M (p^0_k \varphi^0_k+p^1_k \varphi^1_k)$, and using Lemma \ref{initk} with $\xi={\omega}/({\gamma+\omega})$, we get
\bea \bra{\varphi^1_0}e^{t\cL^{\rm ini}}(\rho_D)\ket{\varphi^1_0}&\leq&\sum_{k=0}^M p^1_k \xi^k\\
&\leq& \sum_{k=0}^{M(1-c)} p^1_k \xi^k+\sum_{k=M(1-\varepsilon)+1}^M \xi^k\\
&\leq& \sum_{k=0}^{M(1-c)} (1-\delta)^{c M-k} \xi^k+\sum_{k=M(1-c)+1}^M \xi^k\\
&\leq& M e^{-\mu M},\eea where $\mu$ is the smallest (in absolute value) of all the exponents occurring in the sum. Clearly, every exponent is negative and proportional to $M$. Finally, we can write the whole expression as 
\be 
	\bra{1_c}\rho_c(t)\ket{1_c}\leq M(Ce^{-t\omega}+e^{-\mu M}).
\ee 
Choosing $t\geq \log{(MC/\varepsilon)}/\omega$ then completes the proof.\qed}

One of the crucial properties of the initializer gadget, which is not transparent from Theorem 1, is that the gadget essentially stops acting on the central qubit after a time of order $\log{M}/\omega$. See the final comments for more on this issue.

\begin{lemma}\label{initk}
Let $\varphi^{(0,1)}_0$ and $\cL^{\rm ini}$ be defined as earlier, then 
\be 
	\bra{\varphi^1_0}e^{t \cL^{\rm ini}}(\varphi^1_k)\ket{\varphi^1_0} =
	(1-e^{-t(\omega+\Gamma)})^k\left(\frac{\omega}{\omega+	\Gamma}\right)^k.
\ee
\end{lemma}

\proof{
To see this, consider the action of the initializer channel $e^{t\cL^{\rm ini}}$, where $\cL^{\rm ini}=\cL^{\rm ad}+\cL^{\rm cp}$, on the pure product input state $\varphi^1_k$. 
It is easy to see that 
\bea 
	\cL^{\rm cp}(\varphi^1_k)&=&k\gamma (\varphi^0_k - \varphi^1_k), 
	~~~~~~~~~ \cL^{\rm cp}(\varphi^0_k)=0\label{eqn_Lc},\\
	\cL^{\rm ad}(\varphi^1_k)&=&k\omega(\varphi^1_{k-1}-\varphi^1_k), 
	~~~~~ \cL^{\rm ad}(\varphi^0_k)=k\omega(\varphi^0_{k-1}-\varphi^0_k).
\label{eqn_Ld}\eea
We want to know what the overlap with the stationary state $\varphi^1_0$ is after a time $t$. It is clear from the above equations that once the central  qubit is in $\ket{0_c}$, it never exits it. Since such states have zero overlap with $\rho^1_{\rm ss}$, we can restrict ourselves to dynamics involving states with control bit $\ket{1_c}$. The reduced dynamics read
\be 
	\cL^{\rm ini}(\varphi^1_k)=-k((\Gamma+\omega)\varphi^1_k-\omega \varphi^1_{k-1}).
\ee
For simplicity, we assume that $\omega=\Gamma$. The proof carries through almost unchanged when $\Gamma\neq\omega$. In this case the dynamics further reduce to 
\be
	\cL^{\rm ini}(\varphi^1_k)=-\Gamma k(2\varphi^1_k-\varphi^1_{k-1})\label{rec_relation_L}.
\ee
Now, as we iterate $\cL^{\rm ini}$, we get an expansion in the basis $\{\varphi^1_k\}$,
\be (\cL^{\rm ini})^m(\varphi^1_k)=(-\Gamma)^m\sum_{j=0}^m f_m^k(j)\varphi^1_j,\ee 
where $f^k_m(j)$ are the coefficients of the expansion. It is clear by construction that $f^k_m(j)=0$ if $m+j<k$, and  that they obey the following recurrence relation
\be f^k_{m+1}(j)=-(j+1)f^k_m(j+1)+2jf^k_m(j)\label{rec_relation}.
\ee
The coefficients $f^k_m(j)$ can be calculated explicitly, and are seen to be
\bea f^k_m(j)&=&\frac{1}{2^k}\sum_{l=0}^k\binom{k}{l}(-1)^l(2l)^{m+1}\frac{(l-1)!}{(l-j)!}\frac{2^{j-1}}{j!}, ~~ {\rm for} ~j\geq 1, \label{fkmj}\\
f^k_m(0)&=&\frac{1}{2^k}\sum_{l=0}^k\binom{k}{l}(-1)^l(2l)^{m}, ~~~~~~~~~~~~~~~~~~~~~~~~~~~~ {\rm for} ~ j=0\label{fkm0}.\eea 
Now, we would like to know the overlap with $\varphi^1_0$ of the time evolved state $\varphi^1_k$. This is given by
\bea 
	\bra{\varphi^1_0}e^{t\cL^{\rm ini}}(\varphi^1_k)\ket{\varphi^1_0}&=&\sum_{m=0}^\infty \frac{t^m}{m!}\bra{\varphi^1_0}(\cL^{\rm ini})^m(\varphi^1_k)
	\ket{\varphi_0}\\
	&=&\sum_{m=0}^\infty \frac{(-\Gamma t)^m}{m!}f^k_m(0)\\
	&=& \frac{1}{2^k}\sum_{m=0}^k \frac{(-\Gamma t)^m}{m!} \sum_{l=0}^k \binom{k}{l}(-1)^l(2 l)^m\\
&=& \frac{1}{2^k}\left( 1-e^{-2\Gamma t}\right)^k.
\eea 
By proceeding in exactly the same manner, we get for $\omega\neq\Gamma$ that
\be 
	\bra{\varphi^1_0}e^{t\cL}(\varphi_k)\ket{\varphi^1_0} = 
	\biggl(\frac{\omega(1-e^{-(\Gamma +\omega) t})}{\omega+\Gamma}\biggr)^k.
\ee
\qed}
\begin{lemma}$\label{Lem:added}\|\big(e^{t \cL^{\rm ini}}-T_\infty\big)(\rho_D)\|_1\leq 3 M e^{-t\omega}$.
\end{lemma}
\proof{
We will regard any map as one acting on the abelian algebra in the range of $\cP$. If we define $\eta(T):=\sup_\rho \|T(\rho)-T_\infty(\rho)\|_1$ where the supremum is taken over all density matrices, then our task is to bound $\eta\big(e^{t\cL^{\rm ini}}\big)$. 
To this end note that we can write $\cL^{\rm ini}=\sum_{j=1}^M\cL^{\rm ini}_j$ where $[\cL^{\rm ini}_k,\cL^{\rm ini}_j]=0$ if each $\cL^{\rm ini}_j=\cL^{\rm cp}_j+\cL^{\rm ad}_j$ acts non-trivially only on the central and the $j$'th qubit. Since the Liouvillian is now a sum of  commuting terms we can exploit Thm.\ 6 from Ref.\ \cite{cutoffpaper} which gives
\be
\eta\big(e^{t\cL^{\rm ini}}\big)\leq \sum_{j=1}^M \eta\big(e^{t \cL^{\rm ini}_j}\big)=M\eta\big(e^{t\cL^{\rm ini}_1}\big).
\ee
$\cL^{\rm ini}_1$ acts non-trivially only on two qubits, i.e., it is of the form $\cL^{\rm ini}_1=\Lambda\otimes\id$, where $\Lambda$ is a two-qubit Liouvillian. Now we use that $\eta\big(e^{t\Lambda}\otimes\id\big)=\eta\big(e^{t\Lambda}\big)$.
In order to compute the latter, we realize that $e^{t\Lambda}$ can be represented as a $4\times 4$ stochastic matrix, for which $\eta$ can be computed using a computer algebra program which yields 
\be
\eta\big(e^{t\Lambda}\big)=e^{-t\omega}\big(2+e^{-t\Gamma}\omega/(\omega+\Gamma)\big)\leq 3e^{-t\omega}.
\ee
\qed}

\subsection*{Physically reasonable input state for the initialization gadget}

For completeness, we provide a wide class of physically reasonable input states which satisfy the conditions of Theorem 1, and explicitly give their scaling.
Let $\rho_{\rm in}$ be the (classical) input state, and write it as $\rho_{\rm in} = \sum_{k=1}^N p_k \varphi^1_k$. We assume that $p_k$ follows a truncated discrete normal distribution with mean $\mu$ and variance $\sigma^2$. We further assume that there exist strictly positive constants $\alpha,\beta>0$ with $\alpha<1$ such that $\mu=\alpha M$ and $\sigma^2=\beta M$. For the uniform input state $\1/2^N$, $\{p_k\}$ would indeed be a binomial distribution. We will now take the limit as $N$ goes to infinity and work in the continuum. Hence we assume that $\{p_k\}$ obeys a truncated continuous normal distribution with mean $\mu$ and variance $\sigma^2$. The error in our calculation due to this assumption can be seen to decrease exponentially with $N$.  We wish to evaluate
\bea 
	&&\lim_{t\rightarrow \infty}\bra{\varphi^1_0}e^{t\cL^{\rm ini}}(\rho_{\rm in})\ket{\varphi^1_0} =\\
	&& \lim_{t\rightarrow \infty} \sum_{k=0}^N p_k \biggl(\frac{1-e^{-(\Gamma +\omega) t}}{1+\Gamma/\omega}\biggr)^k \approx\\
	&& \int_0^N  \phi^{\rm tr}_{0,N}\bigl(\frac{x-\mu}{\sigma}\bigr)\xi^{-x}dx,
	\eea
	where $\xi:=1+\Gamma/\omega$ and $	\phi^{\rm tr}_{a,b}$ is the truncated normal probability distribution function restricted to the interval $-\infty\leq a <b
	\leq \infty$ given by
\be \phi^{\rm tr}_{a,b}(\frac{x-\mu}{\sigma})=\frac{\frac{1}{\sigma} \phi(\frac{x-\mu}{\sigma})}{\Phi(\frac{b-\mu}{\sigma})-\Phi(\frac{a-\mu}{\sigma})}.\ee Here $\phi$ is the standard normal distribution function and $\Phi$ is the cumulative normal distribution function. We show the following:
\begin{lemma}
Let $\alpha,\beta>0$ with $\alpha\in(0,1]$. If $\mu=\alpha N$ and $\sigma^2=\beta N$, then 
\be \int_0^{N}\phi^{\rm tr}_{0,N}\biggl(\frac{x-\mu}{\sigma}\biggr)\xi^x dx<Ce^{-N b}\ee for some $N$ independent positive constants $C,b$. 
\end{lemma}
\proof{
We start by bounding the integral expression, and leave the denominator in the truncated normal distribution for later.
Let $z_1:= N(\alpha-\beta \log{\xi})$ and $z_2:= N(2\alpha-\beta \log{\xi})$. Then, a simple calculation shows that 
\be \int_0^{N}dx\phi\bigl(\frac{x-\mu}{\sigma}\bigr)\xi^x = e^{-\frac{\log{\xi}}{2}z_2}\int_0^{N}dxe^{-\frac{1}{2 \beta N}(x-z_1)^2}.\ee
The integral on the right hand side is bounded above by 1, so if $z_2>0$, then the whole expression is bounded above by a function which is decreasing exponentially in $N$. Therefore, we can assume that $z_2<0$ but this implies that $z_1<0$ because $z_1<z_2$. 
Now, note that one can easily bound the complementary cumulative normal distribution from above by observing that for $c>0$, $c<x$ for any $x\in[c,\infty]$, we get
\bea 
	\int_c^\infty e^{-x^2/2}dx&\leq& \int_c^\infty \frac{x}{c}e^{-x^2/2}dx\\
	&=& \frac{1}{c}e^{-c^2/2}.
\eea
We use this to bound the integral as
\bea 
	\int_0^{N}dx\phi\bigl(\frac{x-\mu}{\sigma}\bigr)\xi^x &\leq& 
	e^{-\frac{\log{\xi}}{2}z_2}\int_{-z_1}^{\infty}dx e^{-\frac{x^2}{2\beta N}}\\
	&=& \frac{\sqrt{\beta N}}{z_1}e^{-\frac{\log{\xi}}{2}z_2-\frac{z^2_1}{2 \beta N}}\\
	&=& \frac{\sqrt{\beta }}{\sqrt{N}(\alpha-\beta\log{\xi})}e^{-\frac{N\alpha^2}{2\beta}}.
\eea
Hence, for any nonzero $\alpha$, this expression decreases exponentially with $N$. The cumulative distribution functions in the denominator of the truncated normal distribution function can be bounded in the same way to show that the denominator is 
exponentially close to $1$. 
Thus putting all of the pieces together and bounding the pre-factor (which is strictly decreasing with $N$), we get the desired result. \qed}

\subsection*{Proof of Theorem \ref{cutoffProp}}
\proof{
A simple calculation shows that 
\be e^{t\cL^{\rm cut}}(\phi_0)=e^{-t\gamma}\sum_{k=0}^{N-2}\frac{(t \gamma)^k}{k!}\phi_k + f(t)\ket{0,\dots, 0}
\bra{0,\dots, 0}\ee where $\phi_k$ is equal to the pure state with the first $k+1$ qubits in the $\ket{0}$ state  and the rest of the qubits in the $\ket{1}$ state, and $f(t)$ is some function of $t$. 
Then, since $\tr[e^{t\cL^{\rm cut}}(\rho)]=1$ for any $\rho$, and $e^{t\cL^{\rm cut}}$ is trace and separability preserving, we get that 
\be
	\frac{1}{2}\|\tr_{N-1}[e^{t\cL^{\rm cut}}(\phi_0)]-\ket{0_N}\bra{0_N}\|_1=1-\bra{0_N}\tr_{N-1}[e^{t\cL^{\rm cut}}
	(\phi_0)]\ket{0_N}
\ee and the expectation value can be evaluated as
\bea \bra{0_N}\tr_{N-1}[e^{t\cL^{\rm cut}}(\phi_0)]\ket{0_N}&=&1- \bra{1_N}\tr_{N-1}[e^{t\cL^{\rm cut}}(\phi_0)]\ket{1_N}\\
&=& 1 - e^{-t\gamma}\sum_{k=0}^{N-2}\frac{(t\gamma)^k}{k!}\\
&=& 1 - \frac{\Gamma(N-1,t\gamma)}{\Gamma(N-1)}\\
&=& \frac{\gamma(N-1,t\gamma)}{(N-2)!},\eea where 
\begin{equation}
\Gamma(y)=\int_0^\infty r^{y-1}e^{-r}dr
\end{equation}
is the Gamma function evaluated at $y$, $\Gamma(y,s)=\int_s^\infty r^{y-1}e^{-r}dr$ corresponds to the upper incomplete Gamma function, and $\gamma(y,s)=\int_0^s r^{y-1}e^{-r}dr$  the lower incomplete Gamma function. 
The following relationships holds, 
\begin{equation}
	\gamma(y,s)+\Gamma(y,s)=\Gamma(y), 
\end{equation}
and $\Gamma(n)=(n-1)!$ when $n$ is an integer.
Now, it was shown in Refs.\ \cite{IncomplGamma,IncomplGamma2} that 
\be 
	\gamma(N,N+x\sqrt{N})=(N-1)!(\Phi(x)+\theta(1/\sqrt{N})).
\ee
Thus, in the limit of large $N$, we get the desired result. 
\qed}

Theorem \ref{cutoffProp} gives an estimate of the overlap with the incomplete gamma function near $x=0$ as a function of $N$. In fact, the error term ($\theta(1/\sqrt{N})$) is even exponentially suppressed in $x$. In particular, we treat the case when $x$ is of order $\sqrt{N}$ below, where we consider successive triggering of dissipative operations. 

\subsection*{Imperfect initialization and error analysis}

The next natural question to ask is how much the timer is damaged by imperfect initializations? 
Given that the input state of the timer gadget is initialized with the initialization gadget, we can assume that it will be at least as close in trace norm to the desired input state
\be 
	\phi(\varepsilon)=(\ket{0}\bra{0}(1-\varepsilon)+\ket{1}\bra{1}\varepsilon)\otimes(\ket{1}\bra{1}(1-\varepsilon)+	
	\ket{0}\bra{0}\varepsilon)^{\otimes N-1},
\ee for some worst case error $\varepsilon$ that was shown to scale exponentially in the number of qubits $(\varepsilon \propto e^{-\mu N})$ for some real $\mu>0$. We focus on a single worst case error $\varepsilon$ for 
simplicity of notation; it should be clear that a product input with different site-dependent error can be 
accommodated as well, which will not lead to significantly improved bounds, however. 
Let $\phi:= \ket{0,1,1,\dots,1}\bra{0,1,1,\dots,1}$, 
and let $\phi^{(k)}$ denote $\phi$ but with the bit in the $k+1$ position flipped. 
Then, to first order in $\varepsilon$,
\be
 	e^{t\cL^{\rm cut}}(\phi(\varepsilon))=e^{t\cL^{\rm cut}}[(1-\varepsilon)^N \phi + \varepsilon \sum_{k=0}^{N-1} \phi^{(k)}] 
	+\cO(\varepsilon^2).
 \ee 
By the triangle inequality, and naively bounding the terms involved, 
\bea  
	\bra{0_N}\tr_{N-1}[e^{t\cL^{\rm cut}}(\phi(\varepsilon))]\ket{0_N} &\approx& \bra{0_N}\tr_{N-1}[e^{t\cL^{\rm cut}}(\phi)]	\ket{0_N} \\ &&+N \varepsilon+\cO(\varepsilon^2).
\eea
Hence, the behavior of the trigger is essentially unchanged as long as $\varepsilon$ is exponentially small in the system size.

\bigskip

We now want to consider how the triggering error behaves under the {\it concatenation of timer gadgets}. For this we will invoke a variant of Theorem \ref{cutoffProp}. What we want to show is that the concatenated triggering error is still negligibly small if we take $N$ to be large. Assume perfect initialization, knowing from the previous section that imperfect initialization only adds an error which is linear in $\varepsilon$. Suppose we want to trigger $L$ successive dissipative operations, where $L$ is ${\rm poly}(N)$. Each time step will be provided by $N$ linear timer gadgets. Then, for $l=1,\dots,L$, we want to estimate the probability for operation $l$ to be triggered in a time between $t_1(l)=N(l-1/2)/\gamma$ and $t_2(l)=N(l+1/2)/\gamma$. As can be seen from the last step in the proof of Theorem 2, the probability for correctly triggering an operation between $t_1(l)$ and $t_2(l)$ is one minus the probability of triggering before $t_1(l)$ or after $t_2(l)$. This corresponds to
\be 
	\frac{\gamma(N l,N(l-1/2))}{\Gamma(Nl)}+\frac{\Gamma(N l,N(l+1/2))}{\Gamma(Nl)}.
\ee
These terms can be bounded as follows: It has been shown in Ref.\ \cite{Tricomi} that 
\be 
	\Gamma(a,x)<\frac{e^{-x}x^a}{x-a+1}.
\ee
Then, using Stirling's approximation, and dropping terms which are linear in $N$, we get
\be \frac{\Gamma(N l,N(l+1/2))}{\Gamma(Nl)}\leq e^{-\alpha(l)N}{\rm poly}(N,l),\ee where the degree of the polynomial is fixed, and $\alpha(l)$ is a strictly positive function of $l$. Thus, if we choose $N$ polynomially larger than $L$, the probability of triggering an operation after $t_2(l)$ is exponentially small. Similarly, using the expansion 
\be \frac{\gamma(a,x)}{\Gamma(a)}=e^{-x}\sum_{k=0}^{a-1}\frac{x^k}{k!}, \ee we can bound the other term as
\be \frac{\gamma(N l,N(l-1/2))}{\Gamma(Nl)}\leq e^{-\beta(l)N}{\rm poly}(N,l).\ee
For completeness, we give the functions $\alpha(l)$ and $\beta(l)$,
\bea 
	\alpha(l)&=&(1/2-l(\log{(l+1/2)}-\log{(l)}))/\gamma,\\
	\beta(l)&=&-(1/2+l(\log{(l-1/2)}-\log{(l)}))/\gamma.
\eea 
Thus with $N$ polynomially large in $L$, we can always guarantee that the probability for incorrectly triggering any of the $L$ operations is exponentially small.


\subsection{Dissipative state transfer by measurement}

\begin{figure}[th]
\includegraphics[scale=0.30]{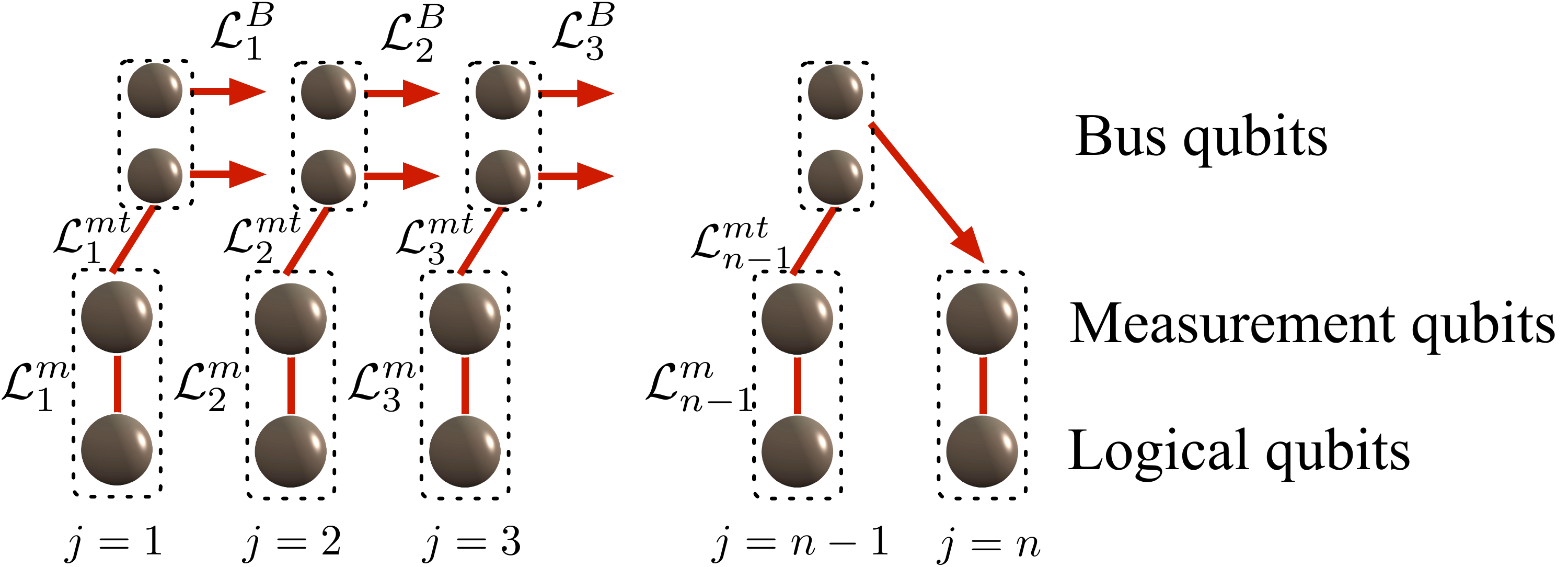}
\caption{Depiction of dissipative state transfer of a qubit state along a chain of odd-$n$ qubits. The lowest line of qubits represents the logical qubits prepared in a cluster state with the 
input state vector $\ket{\varphi_{\rm in}}$ encoded at the first site (see text). The red lines represent ``dissipative measurements'', whereas the red arrows depict dissipative classical information transfer along the bus, and (classical) dissipative binary logical operations from the auxiliary measurement qubits to the bus. The operations are performed from left to right sequentially in time}
\end{figure}

We describe here the general dissipative state transfer protocol which will be slightly more involved than the three qubit chain case described in the main text, in that we need an extra register of qubit to keep track of the ${\rm mod}(2)$ sum of measurement outcomes corresponding to $U^{\rm st}_\Sigma$. The setup consists of odd-$n$ logical qubits initialized in the cluster state vector $\ket{\varphi_{\rm in}}_\cC$, represented  in Fig.\ 2. Associated to the $n$ logical qubits are $n$ measurement qubits, and a classical information bus consisting of $2(n-1)$ ``bus qubits''. The idea underlining the $n$ qubit chain state transfer is the same as for the three qubit chain in the main text, except that the recovery operation
will now be conditioned on the last two bus qubits only. This allows the recovery operation to remain local. Recall that, $U^{\rm st}_\Sigma=\sigma_x^{\gamma_x}\sigma_z^{\gamma_z}$, where $\gamma_x=s_1+s_2+\dots+s_{n-2}$ and $\gamma_z=s_2+s_4+\dots+s_n-1$ and addition is understood as being ${\rm mod}(2)$. The purpose of the classical information bus is to keep track of $\gamma_x,\gamma_z$ as the state transfer moves forward. 

To write out the Lindblad operators, we label the logical qubits with a subscript $l$, the measurement qubits with a subscript $m$ and the bus qubits with a subscript $b_{1,2}$.  The Lindblad operators performing the dissipative measurements are for $j=\{1,\dots, n-1\}$
\bea 
	L^{m}_{1,j}&=&\sqrt{\omega}\ket{+_j}_l\bra{+_j}_l\otimes\ket{0_j}_m\bra{1_j}_m,\\
	L^{m}_{3,j}&=& \sqrt{\omega}\ket{-_j}_l\bra{-_j}_l\otimes\ket{1_j}_m\bra{0_j}_m.
\eea
The Lindblad operators which dissipatively update the bus qubits are for $j=2,\dots,n-1$
\be 
	L^{mt}_j=\sqrt{\omega}\ket{0_j}_m\bra{1_j}_m\otimes(\sigma_j^x)_{b_1}
\ee for $j$ odd, and 
\be 
	L^{mt}_j=\sqrt{\omega}\ket{0_j}_m\bra{1_j}_m\otimes(\sigma_j^x)_{b_2}
\ee for $j$ even.
Next, the information transfer from one state of the bus to the next is given by the following Lindblad operators for $j=\{2,\dots,n-1\}$,
\bea
L^{b_1}_{1,j}=\sqrt{\omega}\ket{0_j,0_j}_{b_1}\bra{0_j,1_j}_{b_1},\\
L^{b_1}_{2,j}=\sqrt{\omega}\ket{0_j,1_j}_{b_1}\bra{1_j,0_j}_{b_1},
\eea 
and similarly for the second line of bus qubits.
Finally, the recovery operation is almost identical to the three qubit chain:
\bea 
	L^B_1&=& \sqrt{\omega}\sigma^z_{j-1}\otimes\ket{0_{j-2},0_{j-2}}_b\bra{0_{j-2},1_{j-2}}_b,\\
	L^B_2&=& \sqrt{\omega}\sigma^x_{j-1}\otimes\ket{0_{j-2},0_{j-2}}_b\bra{1_{j-2},0_{j-2}}_b,\\
	L^B_3&=& \sqrt{\omega}\sigma^z_{j-1}\sigma^x_{j-1}\otimes\ket{0_{j-2},0_{j-2}}_b\bra{1_{j-2},1_{j-2}}_b.
\eea
The $n-1$ measurements will project the system onto a superposition four sets of states. When one traces out the final logical qubit, the first element in the superposition will be $\ket{\varphi_{\rm in}}$, the second one $\sigma_z\ket{\varphi_{\rm in}}$, the third one $\sigma_x\ket{\varphi_{\rm in}}$, and the fourth one $\sigma_x\sigma_z\ket{\varphi_{\rm in}}$. The recovery operation then undoes the bit and phase flips to yield the desired final state.
It is important to note that the bus update operations cannot all be performed simultaneously. So each step $j$ must be performed successively. As the number of steps is polynomial, the scheme remains efficient.


\subsection{ Dissipative quantum error correction} 
In very much the same spirit as for D1WQC, we can consider dissipative error correction models by simulating the known active error correction models.
A promising approach is to perform dissipative measurements on the qubits and process the information on a classical information bus in parallel before applying a corrective operation. The measurements, as well as the corrective operations, can be applied sequentially using a timer gadget, and the encoding and decoding can be performed efficiently with the initialization trick. In this way, it is tempting to think that one can implement Kitaev's toric code \cite{KitaevMem} dissipatively, hence yielding an efficient passive topological memory in 3D. However, a number of obstacles prevent us from certifying that this approach will surely work. First of all, despite a recent promising attempt \cite{toricEC}, there does not exist any protocol for actively correcting Kitaev's 2D toric code with fully local measurements and local classical information processing of polynomial depth. It is even questionable whether such an algorithm can exist in principle. A further obstacle standing in the way of a fault-tolerant dissipative quantum memory is that the gadgets which we introduced can also suffer errors. One way of overcoming this problem is by noting that the dynamics of the gadgets is essentially classical, and hence can be protected using classical ``dissipative" means. The prototypical example of such a classical dissipative error correction is Toom's rule for asynchronous cellular automata, which was proved to scale favorably \cite{asyncToom}. Nevertheless, this introduces a third drawback, which is that of interfacing the fault tolerant classical bits with the qubits; we do not know how to do this.
Thus, although a number of obstacles stand in the way of a passive dissipative quantum memory using the approach outlined above, we nevertheless believe that the flexibility provided by dissipative quantum processing, as illustrated through the timing gadgets, make it a very promising framework for considering passive quantum memories.

\subsection*{Observations}

To conclude, we note two important facts related to the convergence of the schemes discussed here:
\begin{itemize}
\item
[(i)] The timer and initialization gadgets which we have presented and analyzed above have a constant gap, by construction. 
It turns out that in these processes, the second eigenvalue will be highly degenerate, with a degeneracy scaling linearly in $N$. This kind of spectrum is characteristic for processes exhibiting a cutoff phenomenon \cite{Diaconis}.  

This, and the application to D1WQC, implies that universal quantum computing can be performed dissipatively with a global Liouvillian which has a constant gap in the system size.
\item
[(ii)] Once a state has been initialized, the initialization gadget does not completely stop, but in fact keeps acting but with a strength which is exponentially suppressed in $M$. Hence, the action of the initializer after initialization has taken place can be modeled by a quantum channel given by $\tilde{T}=\id\varepsilon+T(1-\varepsilon)$, where $\varepsilon=\cO(e^{-\tau M})$, and $T$ is some noisy channel. Then, if we want to perform operation $S_t$, we can bound the error by 
\be 
	\|\tilde{T}S_t-S_t\|_\infty\leq \varepsilon\|T\|_\infty\leq\varepsilon.
\ee
Hence, the error will be exponentially small. 
An illustrative example of the above statement goes as follows: Consider the rudimentary conditional operation acting on two qubits described by the Lindblad operator $L^{\rm cond}=\sqrt{\gamma}\ket{0}\bra{1}\otimes\ket{1}\bra{1}$, and assume that it is acting on the state $\rho=(1-\varepsilon)\ket{1,0}\bra{1,0}+\varepsilon\ket{1,1}\bra{1,1}$. In other words, the conditioning will be governed by $\varepsilon$. It is easy to see that
\bea 
	e^{t\cL^{\rm cond}}(\rho)&=&\ket{1,0}\bra{1,0}+\varepsilon(\ket{1,0}\bra{1,0}+\ket{0,1}\bra{0,1})\\
	&+& \varepsilon e^{-t	
	\gamma}(\ket{1,1}\bra{1,1}-\ket{0,1}\bra{0,1}).
\eea
Thus, as long as $\varepsilon$ is very small, the conditional preparation will be effectively stopped. In the case of the initialization gadget, this $\varepsilon$ is of order $e^{-t\mu}$. 
\end{itemize}

\end{appendix}


\begin{thebibliography}{30}
\bibitem{commentQDots} 
	However, many physical systems such as quantum dots naturally have a star structure, so that ultimately, the type of 	locality desired for the initialization gadget will depend on the physical setup at hand.
\bibitem{commentRingGadget}
	One particularly appealing one is to consider a ring of $2 N$ qubits with the left half initialized to all $\ket{1}$'s and 	
	the right side initialized to all $\ket{0}$'s. Then apply the following dissipative process to pair of neighboring qubits, 
	flip the right qubit conditional on the left qubit being in a different state than the right one. This will lead to a 
	conserved pattern of one half of consecutive $\ket{0}$'s and one half of consecutive $\ket{1}$'s but which are rotating 
	clockwise at a constant speed. Then picking one particular bit as a control bit, we get a cyclical tic at time $\cO(N)$. 
	Moreover, the cutoff for this process will be exactly the same as in Theorem\ \ref{cutoffProp}, and hence the error 
	analysis also carries through. For processes requiring a regular sequence of dissipative operations this gadget is 
	more efficient than multiple linear gadgets.
\end{thebibliography}
\end{document}